\documentclass[aps,prl,twocolumn,showpacs,preprintnumbers,amsmath,amssymb]{revtex4}

\usepackage{graphicx}
\usepackage{dcolumn}
\usepackage{bm}

\begin{document}


\title{X-ray phase-contrast imaging for laser-induced shock waves}

\author{L. Antonelli}
\email{luca.antonelli@york.ac.uk}
\altaffiliation{York Plasma Institute, Department of Physics, University of York, York, YO10 5DQ, U.K.}
\author{S. Atzeni}%
\author{A. Schiavi}%
\affiliation{Dipartimento SBAI, Universit\`a degli Studi di Roma "La Sapienza", Via Antonio Scarpa 14, 00161, Roma, Italy}%

\author{F. Barbato}
\author{D. Bleiner}
\affiliation{Empa - Swiss Federal Laboratories for Materials Science and Technology, \"{U}berlandstrasse 129, CH8600 D\"{u}bendorf, Switzerland}%

\author{D. Batani}
\author{G. Boutoux}
\author{D. Mancelli}
\author{J. Trela}
\affiliation{Universit\'e de Bordeaux, CNRS, CEA, CELIA (Centre Lasers Intenses et Applications), UMR 5107, F-33405 Talence, France}%

\author{L. Volpe}
\author{G. Zeraouli}
\affiliation{CLPU, Centro de L\'{a}seres Pulsados, Edificio M5. Parque Cient\'{i}fico. C/ Adaja, 8. 37185 Villamayor - Salamanca - Spain}%
\altaffiliation{Universidad de Salamanca Facultad de Ciencias. Plaza de los Ca\'{i}dos, s/n. 37008 Salamanca - Spain}
\author{P. Bradford}
\author{N. Woolsey}
\affiliation{York Plasma Institute, Department of Physics, University of York, York, YO10 5DQ, U.K.}%

\author{V. Bagnoud}
\author{C. Brabetz}
\author{P. Neumayer}
\author{B. Zielbauer}
\affiliation{GSI Helmholtzzentrum f\"{u}r Schwerionenforschung GmbH, Planckstra\ss{}e 1, 64291 Darmstadt, Germany}%

\author{B. Borm}
\affiliation{Goethe-Universit\"at Frankfurt,
Max-von-Laue-Stra\ss{}e 1, 60438 Frankfurt am Main, Germany}%

\date{\today}

\begin{abstract}
X-ray phase-contrast imaging (XPCI) is a versatile technique with wide-ranging applications, particularly in the fields of biology and medicine. Where X-ray absorption radiography requires high density ratios for effective imaging, XPCI is more sensitive to the density gradients inside a material. In this letter, we apply XPCI to the study of laser-driven shockc waves. We used two laser beams from the Petawatt High-Energy Laser for Heavy Ion EXperiments (PHELIX) at GSI: one to launch a shock wave and the other to generate an X-ray source for XPCI. Our results suggest that this technique is suitable for the study of warm dense matter (WDM), inertial confinement fusion (ICF) and laboratory astrophysics.
\end{abstract}

\pacs{Valid PACS appear here}
\maketitle

X-ray phase contrast imaging (XPCI) is based on the phase-shift of X-ray photons induced by a density gradient. In the presence of a strong density variation, coherent radiation will be deflected from regions of higher density to regions of lower density. If we consider an XPCI image of a solid-vacuum interface, one would observe a bright border around the interface and a dark border inside the solid. The actual position of the interface will be located between the areas of maximum and minimum intensity.
Synchrotrons and Free Electron Lasers are ideal experimental platforms for XPCI because they can deliver coherent radiation at high energy (to limit absorption) and high flux \cite{schropp2015,hawreliak2017,nagler2017}. It is also possible to use broadband incoherent radiation for phase contrast imaging, however the corresponding X-ray source must be very small. One method for generating small-scale X-ray sources suitable for XPCI is to use laser-irradiated solid targets \cite{Fourmaux2011,kneip2011}. XPCI has already made an important contribution to the fields of biology and medicine \cite{davis1995,snigirev1995,momose1996,wilkins1996}, but laser-driven XPCI could also be applied to studies of warm dense matter (WDM), laboratory astrophysics and inertial confinement fusion (ICF). Large-scale laser facilities such as the National Ignition Facility (NIF)\cite{miller2004} and Laser MegaJoule (LMJ)\cite{fleurot2005} enable us to study matter in extreme conditions and both have dedicated beamlines for target probing: the Advanced Radiographic Capability (ARC) \cite{crane2010} and the PETawatt Aquitaine Laser (PETAL) \cite{batani2013}. With the increased precision and detail available through XPCI, the development of XPCI lines on these facilities could open up new possibilities in diagnostic imaging.

Though laser-driven X-ray absorption radiography has been successfully demonstrated on many experiments (some examples are reported in \cite{benuzzi2006,lepape2008,antonelli2017,kritcher2014,delsorbo2015}), XPCI using laser-produced X-ray sources has been less intensively studied. A significant advance in the application of laser-driven bremsstrahlung X-ray sources to XPCI was shown by Workman et al. in 2010 \cite{workman2010}, however the quality of the images they obtained did not allow for a comprehensive study of shock wave characteristics. In \cite{montgomery2004}, a numerical study of cryogenic beryllium capsules using phase-contrast imaging is presented, however a proof-of-principle laser experiment is necessary to pin down the requirements of a single-shot, laser-produced X-ray source for XPCI.
In this letter, we present results from an XPCI experiment performed at the PHELIX facility \cite{neumayer2005}. The experiment was divided into two parts: the characterization and optimization of XPCI using a static object and the application of this diagnostic to the study of laser-driven shock waves. This proof-of-principle experiment shows that XPCI is a powerful tool to study shock waves due to its high sensitivity to density gradients. The total energy delivered by the laser was 50~J, divided equally between the short pulse beam (to generate the X-ray source) and the long pulse beam (to drive the shock wave). Even at these low energies, it was possible to observe details of the shock shape and internal structure, as well as the locations of the shock front and rarefaction wave. Our data is compared with synthetic images obtained from the hydrodynamic simulation code DUED \cite{atzeni2005} coupled with an XPCI simulation tool which solves the Kirchoff-Fresnel equation using the Fresnel approximation \cite{cowley1995}.

\begin{figure}
\includegraphics[scale=0.35]{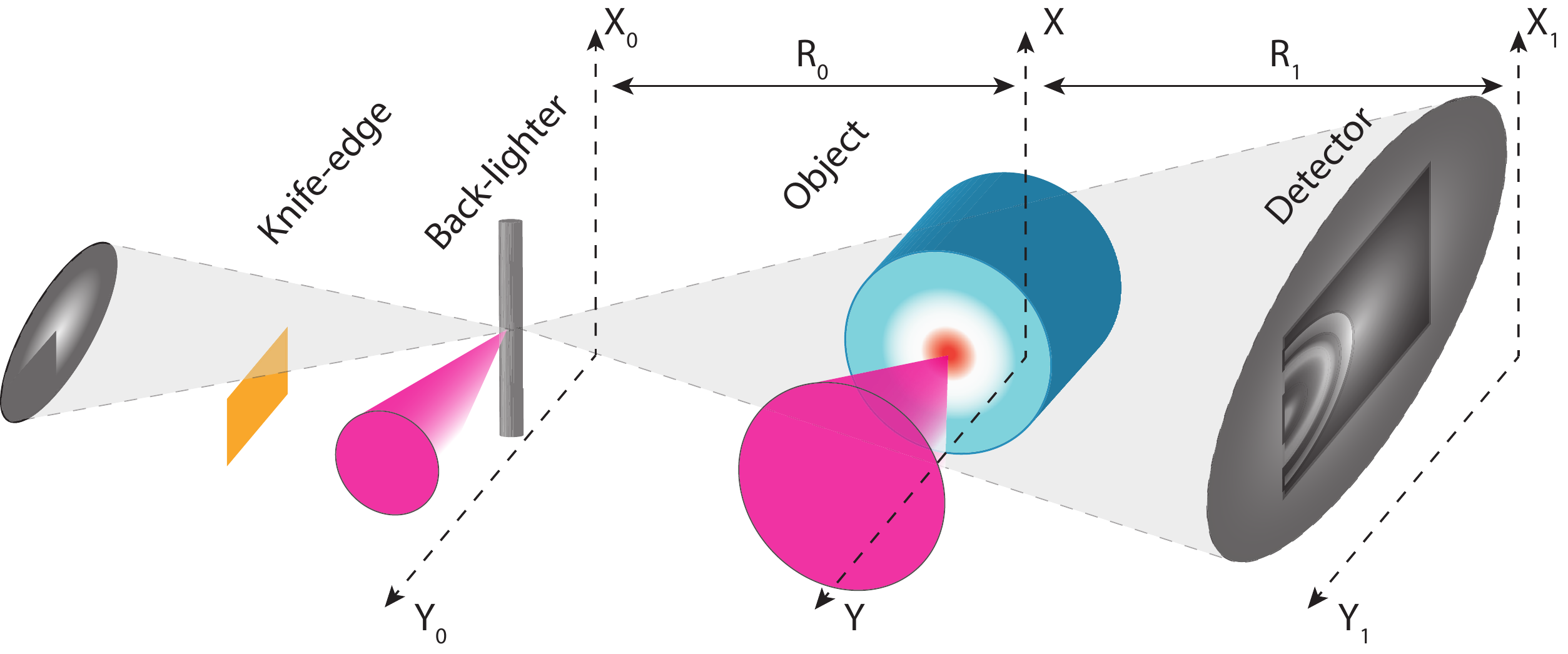}
\caption{\label{setup} Experimental set-up showing the short-pulse backlighter and long-pulse drive beams.}
\end{figure}
PHELIX is a flash lamp-pumped Nd:glass system. It can deliver two laser beam: one short beam with a pulse duration of 0.5 ps and a long beam with a pulse duration between 1 and 10 ns. The maximum energy per beam is 25~J. We irradiated a 5 \textrm{$\mu$m} diameter Tungsten wire with the short pulse to drive a miniature bremsstrahlung broadband X-ray source for XPCI. The long pulse was used to launch a shock wave in a plastic target (\textrm{$C_{8}H_{8}$}). The experimental layout can be seen in Figure \ref{setup}. A knife edge was used to characterize the source dimensions on each shot, while a bremsstrahlung cannon and a highly ordered pyrolytic graphite (HOPG) crystal spectrometer were used to characterize the X-ray spectrum. We used two different detectors to record our images: an Andor CCD camera and an Image Plate (IP). The CCD camera had better spatial resolution, while the IP had better sensitivity \cite{boutoux2015}. In the first phase of our experiment, we took images of static plastic wires to find the best configuration of source size, detector resolution, source-object and object-detector distance. Then, in the second phase of our experiment, we used XPCI to study the propagation of a shock wave in a laser-irradiated polystyrene cylinder with a diameter of 300 $\mu$m.
As mentioned earlier, we used the short pulse beam to generate a bremsstrahlung X-ray source. A 25~J, 0.5 ps laser pulse was focussed onto a tungsten wire with a 5~$\mu$m focal spot, leading to on-target intensities of around $6\times 10^{19}$~Wcm$^{-2}$. Under these conditions, a large portion ($\sim{10-20}\%$) of the incident laser energy is transferred to relativistic electrons \cite{shonelein2016} that propagate through the wire and emit bremsstrahlung radiation. Spectral data from the HOPG crystal spectrometer revealed L-shell emission from 8200 eV to 8400 eV, however the main contribution was at lower energy, as detected by the bremsstrahlung cannon. The source size was measured using a knife edge. In the horizontal direction, the source was measured to be 5~$\mu$m across (the same as the wire diameter),  while the size measured along the wire was 30~$\mu$m. A 500~$\mu$m thick Polymethyl methacrylate (PMMA) window plus 40~$\mu$m thick Al filter were placed in front of our detector. The transmission of these filters was 2$\%$ at 5 keV to 63$\%$ at 10 keV. This choice allows us to remove any contribution coming from the interaction of the long pulse with the target during the shock wave generation. The characteristics of our X-ray source are consistent with phase-contrast enhancement. We can prove this by considering the transversal coherence $l_{t}$, which is the minimum distance between two points in the transverse direction with a correlated phase, defined as:
\begin{equation}
l_{t}\approx\frac{R_{0}\lambda}{s}
\end{equation}
where $R_{0}$ is the distance from source to sample, $s$ is the source size (in our case $l_{t}$ has a different value in the vertical and horizontal directions due to the different source size) and $\lambda$ is the X-ray wavelength.
If we consider Fresnel diffraction, the recorded pattern on the detector surface results from the superposition of waves coming from a coherence area of size $l_{t}$. In other words, $l_{t}$ has to be larger than the scale length for the structure to be resolved. In the case of a laser-induced shock wave, $l_{t}$ should be of the order of a few microns.
The X-ray wavelength ranged from 1.2 to 2.0~\AA. We tested different distances $R_{0}$ throughout the experiment, before fixing it at 27 cm. Taking source dimensions into account, the minimum value of $l_{t}$ is 1~$\mu$m in the vertical direction and its maximum is 10~$\mu$m in the horizontal direction (where the source size is limited to 5~$\mu$m).

\begin{figure}
\includegraphics[scale=0.4]{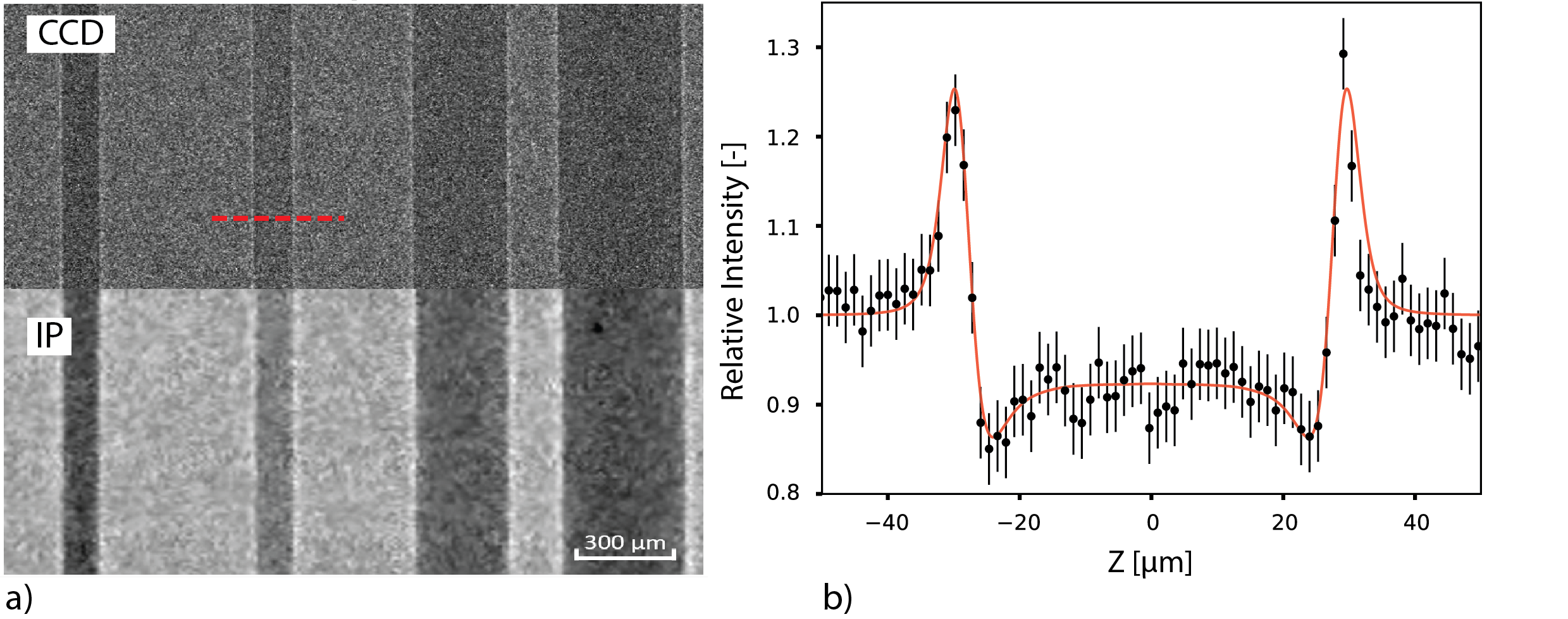}
\caption{\label{14841_image}a) XPCI image of a plastic wires. The upper part correspond to the CCD detector while the lower part to the IP. b) X-ray transmission profile from experiment (black dots) and synthetic profile (red line) corresponding to the second wire from the left (red dashed line).}
\end{figure}
First, we used the short pulse beam on its own to produce an incoherent polychromatic source (bremsstrahlung radiation).
Figure \ref{14841_image}a shows an XPCI image of cylindrical nylon wires with diameters from 120 to 500~$\mu$m. The source-object distance $R_{0}$ was 27~cm while the object-detector distance $R_{1}$ was 94~cm. The choice of these distances give an average value for $l_{t}$ of about 10 $\mu$m. The detector used was an Andor CCD X-ray camera. The image shows the presence of phase contrast at the edges of the wire and low levels of absorption (around 10$\%$ of the incident X-ray radiation below 5 keV).
If we consider the transverse profile of one of the wires with respect to the wire axis (represented by black dots in Figure \ref{14841_image}b), the phase contrast edges are clearly visible while absorption plays a minor role. The red line in Figure \ref{14841_image}b is the synthetic profile calculated using our own code. The code was designed to calculate X-ray absorption and phase contrast, taking the X-ray spectrum, source size and spatial intensity distribution into account. Considering the experimental limitations (detector resolution, low photon flux, etc.) there is good agreement between experiment and simulation.

\begin{figure*}
\includegraphics[scale=0.7]{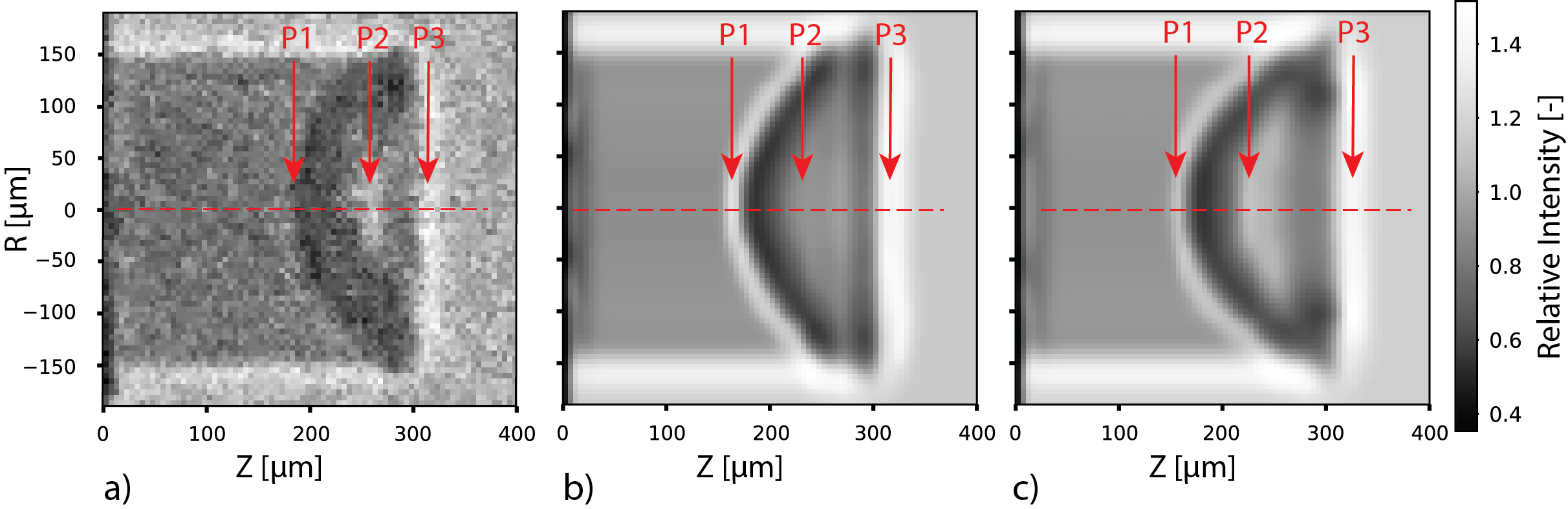}
\caption{\label{comparison_image} Comparison between a) XPCI from the experiment, b) A synthetic XPCI image calculated using a specific module coupled to the DUED hydrodynamic code using the nominal focal spot and c) The synthetic XPCI image, using a small focal spot.}
\end{figure*}

In the second phase of our experiment, we used XPCI to probe a dynamic process: laser-induced shock wave propagation through 300 $\mu$m diameter plastic cylinders. The energy available for the laser beam was 25~J with a focal spot diameter of 50~$\mu$m. We used a source-object distance of $R_{0} =$ 24.5 cm and an object-detector distance of $R_{1} =$ 205.5 cm.
Figure \ref{comparison_image}a shows a laser-driven shock wave propagating inside a plastic cylinder. This image was taken 6 ns after the end of the driving pulse. There is evidence of both absorption and phase-contrast processes, with the strongest phase-contrast at the target-vacuum edge (P3) and inside the shocked region (P2). It is also present on the shock wave front (compressed-uncompressed interface P1). XPCI is sensitive to density variations and can provide information on shock wave propagation even at moderate laser intensity. The X-ray intensity inside the shock wave is higher than the vacuum background intensity, suggesting that a strong density gradient is present in the low-density region before the shock front (P2). To model this internal structure, we ran a number of simulations using the hydrodynamic code DUED coupled to a bespoke XPCI simulation code. Initially, we assumed a super-Gaussian focal intensity distribution with a diameter of 50 $\mu$m, a square time shape with a pulse duration of 2 ns and an energy of 25~J. 
Hydrodynamic simulations performed with the code DUED correctly reproduced the dynamics of the shock wave by assuming that the energy deposited in the target was about half the nominal laser energy. This is consistent with the expected levels of absorption and refraction of the laser beam (taking into account the absence of optical smoothing).
\begin{figure*}
\includegraphics[scale=0.7]{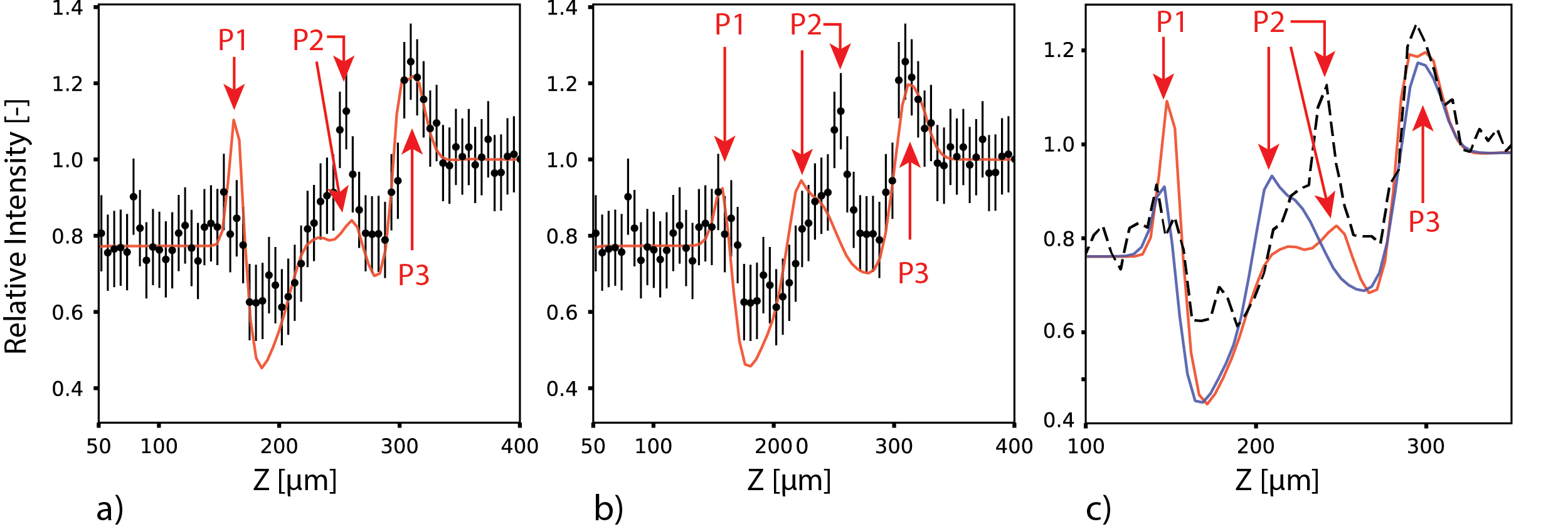}
\caption{\label{comparison_profile} Comparison between the profile along the propagation axis of a) XPCI simulation with nominal focal spot dimension, b) reduced focal spot dimension. Image c) shows the profile comparison between the simulations (red and blue lines) with the experiment (black dashed line).}
\end{figure*}
\begin{figure*}
\includegraphics[scale=0.7]{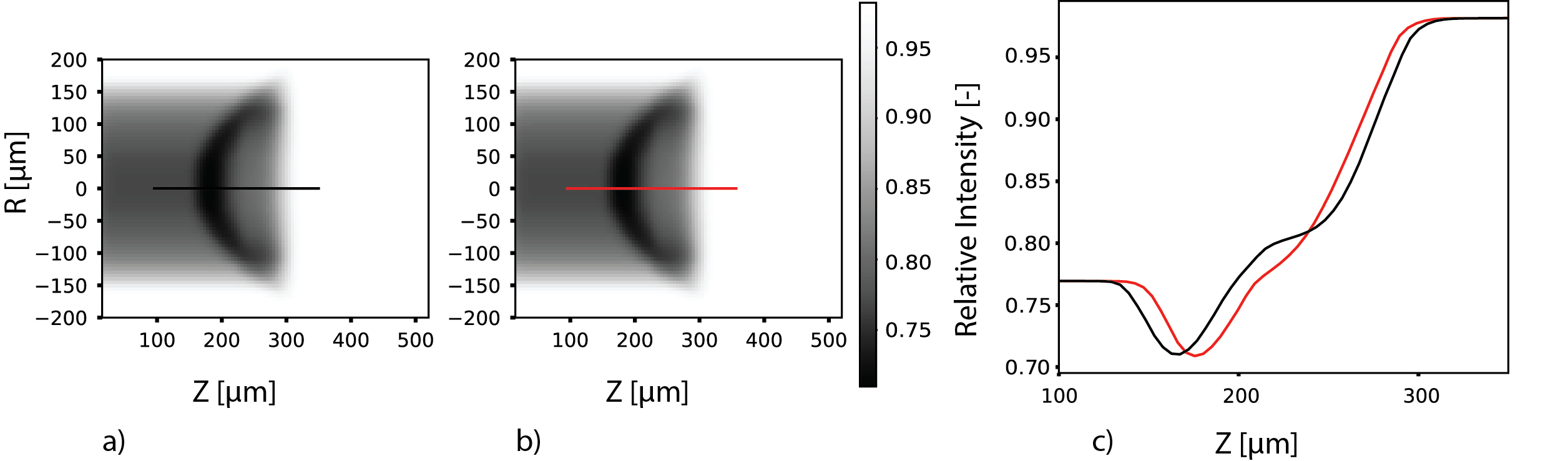}
\caption{\label{comparison_absorption} Synthetic radiographs corresponding to the simulation with a) nominal focal spot and b) reduced focal spot. The image c) compares the intensity profiles along the axis of the image a) (black line) and b) (red line).}
\end{figure*}
Figure \ref{comparison_image}b shows the hydrodynamic simulation with phase-contrast simulation tool. Although reducing the energy in the simulation allowed us to match the position of the wavefront on-axis, the synthetic image looks quite different from the experimental data. Moreover the phase contrast at the simulated shock front is higher than in the experiment, which implies that these simulations have unrealistically steep density gradients. 
Figure \ref{comparison_profile}a shows the intensity profile along the horizontal axis (to help reduce noise, the central line was averaged with the two nearest points in the transverse direction). Our code reproduces the peak corresponding to the vacuum-target interface and also the position of the shock front. The peak intensity is different, however, and we can deduce that the simulation is predicting a higher density-ratio between the shocked and unshocked regions since the width of the simulated and experimental peaks is comparable. Though significant phase-contrast is visible inside the simulated shock, the structure is not well-reproduced. Instead of a single, intense peak, the red profile has a weaker, bimodal structure. In addition, the bright region is much more extended in the simulation than the experiment. One explanation for the discrepancy comes directly from the experimental image: A localized bright region inside the shock wave is indicative of a strong density gradient which will ``deflect" photons from the higher density region to the lower one. This single intensity peak is probably due to the rarefaction wave which stands behind the shock front and inside the shocked material. This strong 2D evolution is more consistent with a smaller focal spot.
We could not characterize the focal spot at full power and there was no phase plate to smooth the focal spot distribution. It is therefore reasonable to expect high intensity spikes which would affect laser energy deposition. In addition, considering the wavelength used, we were also more susceptible to parametric effects which could modify the energy absorption.

To test this hypothesis, we performed several simulations where we progressively reduced the the laser spot size from 50~$\mu$m down to a few microns. Results for the smallest focal spot are detailed in Figure \ref{comparison_image}c. Here, we can distinguish a bright region corresponding to a single phase-contrast peak that is broadly consistent with our experimental results. While the agreement is not perfect, this simulation proves that a spike in laser intensity can dramatically affect the resulting phase-contrast image. The laser energy was kept at 25~J in these simulations.
In Figure \ref{comparison_profile}b, we present on-axis intensity profiles for the experiment alongside numerical simulations with a smaller focal spot. A single, intense peak is apparent in the central region that is qualitatively consistent with the experiment.
One explanation for a smaller focal spot in our experiment could be self-focusing of the laser beam \cite{young1995,young1988}. The laser pulse duration was long ($\tau$~=~2~ns), which would allow the laser to interact with plasma generated earlier in the interaction.
In order to improve the agreement, a more detailed characterization of the focusing condition is required. However, even with this limitation, the obtained XPCI images are high-quality and allow a detailed study of shock wave dynamics.
By contrast, X-ray absorption radiography does not provide us with the same level of detail.
To prove this, we can compare the synthetic absorption radiography of the two simulations.
The results are shown in Figure \ref{comparison_absorption}a and \ref{comparison_absorption}b. It is much harder to identify differences between the simulations using X-ray absorption than with phase contrast imaging (cf. Figure \ref{comparison_image}b and \ref{comparison_image}c).

In Figure \ref{comparison_profile}c, we show on-axis intensity profiles for the images in Figure \ref{comparison_image}b and \ref{comparison_image}c. The red and black absorption profiles are similar, but the red is slightly shifted with the suggestion of a central bump. The energy deposition is different in the two cases and this can cause a difference in the shock velocity. In the case of a small focal spot (red), the 2D effects are stronger and they cause energy to be diffused transversely to the propagation axis. Figure \ref{comparison_profile} shows the same profile with the phase contrast included. Even accounting for the low resolution of the detector (IP), the structure of the shock wave and the rarefaction wave have been successfully observed. Experimental work has already been done to compare absorption radiography and XPCI in other contexts. In \cite{arfelli1998}, for example, X-ray imaging of a locust demonstrates that XPCI is able to detect features that are completely absent from images made using absorption radiography.
The superior sensitivity of XPCI could open up new avenues in the study of warm dense matter, laboratory astrophysics or hydrodynamic instabilities at a large range of densities. Indeed, the ability to image rarefaction waves will allow us to study equations of state in a new way \cite{foord2004}.
In this work we have presented new XPCI data from the PHELIX laser system at GSI in Germany. We have demonstrated XPCI with a laser-produced X-ray broadband source from a single beam. Our set-up was first tested on static objects and then used to image a laser-driven shock wave. In both cases, phase-contrast at the density interfaces could be clearly discerned. The intrinsic sensitivity of XPCI to density gradients enabled us to observe subtle details in the structure of the shock wave. Compared to X-ray absorption radiography, XPCI is more sensitive to density gradients and works well with polychromatic sources. This experiment proved that XPCI can be a useful tool in studies of warm dense matter and high energy density physics, clearing the way for testing on large-scale laser facilities.
\begin{acknowledgments}
This work benefitted from the support of COST Action MP1208 'Developing the physics and the Scientific Community for Inertial Fusion' and by the EUROfusion project "Preparation and Realization of European Shock Ignition Experiments". The research leading to these results has received funding from LASERLAB-EUROPE (grant agreement no. 654148, European Union's Horizon 2020 research and innovation programme).
\end{acknowledgments}
\end{document}